\pdfoutput=1

\documentclass[11pt, a4paper]{article}
\usepackage{jcappub}
\usepackage[footnotesize]{caption}
\usepackage{comment}

\numberwithin{equation}{section} 
\numberwithin{figure}{section}
\numberwithin{table}{section} 
\begin{document}

\begin{titlepage}

\thispagestyle{empty}
\setcounter{page}{0}

\vskip 1.5cm

\begin{center}
{\LARGE\bf An Hamilton-Jacobi formulation\\ of anisotropic inflation}

\vskip 2cm

{\large  Francesco Cicciarella
	, Joel Mabillard${}^{a,}$\footnote{\textbf{Corresponding author:} joel.mabillard@gmail.com},\\ Mauro Pieroni${}^{b,}$\footnote{mauro.pieroni@uam.es}, Angelo Ricciardone${}^{c,}$\footnote{angelo.ricciardone@pd.infn.it}}\\[3mm]
{\it{
${}^{a}$ Higgs Centre for Theoretical Physics, School of Physics and Astronomy, University of Edinburgh, Edinburgh, EH9 3JZ, United Kingdom\\
${}^{b}$ Instituto de F\'isica Te\'orica UAM/CSIC C/ Nicol\'as Cabrera 13-15 Universidad Aut\'onoma de Madrid Cantoblanco, Madrid 28049, Spain\\
}
${}^c$ INFN, Sezione di Padova, via Marzolo 8, I-35131, Padova, Italy}
\end{center}

\vskip 1cm
\centerline{ {\bf Abstract}}{	 

Classifying inflationary scenarios according to their scaling properties is a powerful way to connect theory with observations. A useful tool to make such a classification is the $\beta$-function formalism. By describing inflation in terms of renormalization group equations, within this framework, it is possible to define universality classes, which can be considered as sets of theories that share a common scale invariant limit. In this paper we apply 
the formalism to define such classes of universality for models of inflation where the inflaton is coupled to gauge fields. We show that the formalism may consistently be extended to capture the peculiar features of these models such as statistical anisotropy. We also obtain some consistency conditions which serve as useful guidelines for model building. }

\end{titlepage}
\tableofcontents

\newpage
\section{Introduction}
\label{sec:introduction}

Symmetries have always played an important role in theoretical physics. In the context of early time cosmology, symmetry arguments may provide a powerful guideline to grasp some details on the physics that occurred during inflation. In particular, it is interesting to understand whether they could be used to infer some of the properties of the perturbations imprinted in the Cosmic Microwave Background (CMB). Following this logic, it becomes natural to classify inflationary models on the basis of scaling/symmetry properties in the deep inflationary phase (where the dynamics features an attractor). In this classification scheme, the $\beta$-function formalism~\cite{Binetruy:2014zya} has shown to be a powerful tool. This framework, which is based on the application of Hamilton-Jacobi (HJ) formalism to cosmology~\cite{Salopek:1990jq}, is based on a formal analogy between the Renormalization Group Equation (RGE) in quantum field theory (QFT) and the equations describing the evolution of the inflaton field in a cosmological background. The behavior of the Universe during inflation is well described, in term of the $\beta$-function formalism, as a departure of the corresponding RGE from a de Sitter (dS) fixed point. The characterization of the $\beta$-function in the neighborhood of such a point defines a universality class of models. In this way the plethora of inflationary potentials~\cite{Martin:2013tda}, which are the discriminator of inflationary models, are classified in a small set of classes according to the $\beta$-function behavior close to the dS fixed point. It is worth mentioning that beyond the formal resemblance between inflation and RGE, more solid arguments in support of this analogy arise from the application of holography~\cite{Maldacena:1997re} to cosmological setups~\cite{Skenderis:2006jq, McFadden:2009fg, McFadden:2010na}\footnote{For other application of holographic techniques to cosmology see for example~\cite{McFadden:2010vh, Bzowski:2012ih, Kiritsis:2013gia, Garriga:2014ema, Garriga:2014fda, Afshordi:2016dvb, Kiritsis:2016kog, Afshordi:2017ihr, Conti:2017pqc, Nitti:2017cbu, Bilic:2018ffh, Bilic:2018uqx}.}. \\

Up to now such a formalism has been applied to inflationary scenarios with scalar field(s) evolving in isotropic space-time, typically Friedman-Lemaitre-Robertson-Walker (FRLW), which is in very good agreement with CMB~\cite{Ade:2015xua,Ade:2015lrj, Ade:2015hxq,Kim:2013gka,Ramazanov:2013wea, Ramazanov:2016gjl} and Large Scale Structure (LSS) observations~\cite{Pullen:2010zy, Sugiyama:2017ggb, Soltis:2019ryf}. However, since the role of new degrees of freedom during inflation is an extremely interesting and actual topic~\cite{Drewes:2019rxn, Figueroa:2018twl, Wands:2002bn}, in this paper we discuss its application to scenarios where, besides the inflaton, vector fields are present during inflation. There are many models in the literature involving vector fields:  the first concrete proposal was realized in~\cite{Ford:1989me, Golovnev:2008cf} where inflation was driven by massive vector fields (see also~\cite{Kanno:2008gn}). However, since there must be an equilibrium between the rapid dilution of energy density due to the  expansion of the Universe and the amount of generated anisotropy, the dynamics of vector fields during inflation happens to be non-trivial. Moreover, it was shown in~\cite{Bluhm:2008yt, Himmetoglu:2008zp, Himmetoglu:2008hx} that these two  original models suffer from instabilities. In particular the longitudinal degree of freedom, which appears since the $U(1)$ symmetry is broken, turns out to be a ghost. \\

After the first proposals, several models where the dynamics of the vectors does not present pathology (and energy density and anisotropy are under control) were introduced: for example this can happen when $U(1)$ gauge field(s) are coupled to a scalar inflaton as in~\cite{Watanabe:2009ct}, to an axion as in~\cite{Turner:1987bw, Garretson:1992vt, Anber:2006xt} or with non-Abelian gauge fields~\cite{Maleknejad:2011sq, Maleknejad:2011jw,Maleknejad:2011jr}.  In this work we focus on the first of these cases and in particular we consider two (sub)cases: one where a triad of $U(1)$ gauge fields is evolving isotropizing the Universe and one where the gauge field, evolving in a Bianchi type I universe, leaves particular imprints on the dynamics and on CMB observables. In particular the power spectrum (and the bispectrum) acquires a direction dependent contribution which is a measure of the degree of anisotropy~\cite{Ackerman:2007nb, Dulaney:2010sq, Gumrukcuoglu:2010yc, Watanabe:2010fh, Soda:2012zm, Bartolo:2012sd, Bartolo:2015dga}. \\

In this work we apply the $\beta$-function tools to such scenarios developing a consistent framework for the formalism in non isotropic space-time and we show how to build universality class for this kind of models. We first extend the formalism to what we call `isotropic ansatz', where a triad of gauge fields is evolving in an isotropic FLRW metric, and we extract information on the dynamics studying the space of trajectories in terms of the $\beta$-function.  Then we move to the central part of our paper, where the formalism is extended to the `anisotropic ansatz', where the gauge-field is expanding along one particular direction in a Bianchi type I space-time. In this case we introduce a `second' superpotential (and a new function which we call $\gamma$) to describe the growth of the anisotropies. Such a gauge field configuration leaves very peculiar signatures in the observables both on scalar and tensor perturbations. We show how the constraints, mainly arising from CMB observations, may help to characterize the HJ variables and we finally apply the techniques to some specific models of inflation. \\

The structure of the paper is the following: in Sec.~\ref{sec:model_definition} we present the model and the two ansatz used for the HJ analysis. In Sec.~\ref{sec:beta_function_general} we firstly  review the $\beta$-function formalism and then we apply it to both the isotropic case and the anisotropic case. We then discuss the implications of the formalism on CMB observables like statistical anisotropy in the power spectra. We finally apply the techniques to a specific class of inflationary models.  Our conclusions are given in Sec.~\ref{sec:conclusions}. In the Appendices we report some useful formulae and other concrete applications of our formalism.

\section{Inflation in presence of classical gauge field backgrounds}
\label{sec:model_definition}
Models where there is a coupling between the inflaton and gauge fields have been extensively studied in literature (see for example~\cite{Turner:1987bw, Freese:1990rb, Garretson:1992vt, Anber:2006xt, Watanabe:2009ct,  Maleknejad:2011sq, Maleknejad:2011jw, Maleknejad:2011jr, Soda:2012zm, Maleknejad:2012fw, Pajer:2013fsa, Domcke:2016bkh, Peloso:2016gqs, Domcke:2018rvv}). In particular,  when the coupling balances the dilution due to the expansion, this may result in non-trivial signatures in the primordial fluctuations such as primordial statistical anisotropy or amplification of the scalar and tensor fluctuations. In this work we consider a scalar inflaton $\phi$ coupled to a certain number $\mathcal{N}$ of abelian ($U(1)$) gauge fields $A_\mu^a$ (labeled by the index $a$) described by the action\footnote{The metric has $(-,+,+,+)$ signature, $m_p \simeq 2.4 \cdot 10^{18} $ GeV$/c^2$ denotes the reduced Planck mass (in the following we set $m_p = 1$) and $t$ denotes the cosmic time (conformal time $\tau$ is defined as $a \textrm{d} \tau \equiv  \textrm{d} t$).}:
\begin{equation}
	\label{eq:general_action}
	 \mathcal{S}  = \int \sqrt{|g|} \textrm{d}^4 x \left[  \frac{ m_p^2 R}{2 } - P(\phi, X) - \frac{f(\phi,X)}{4} g^{\mu\nu} g^{\alpha\beta} F^a_{\mu \alpha}F^a_{\nu \beta} \right]  \; ,
\end{equation}
where $F^a_{\mu\nu} \equiv \partial_\mu A^a_\nu-\partial_\nu A^a_\mu$ is the usual (abelian) field strength tensor for the gauge fields, $P(\phi, X)$ and $f(\phi,X)$ are two generic functions of $\phi$ and $X \equiv \frac{1}{2}g^{\mu\nu}\partial_\mu\phi\partial_\nu\phi$. Notice that according to this definition we have $P = - p$, where $p$ is the pressure associated with the scalar field (which in the simplest realization of inflation~\footnote{By `simplest realization of inflation' we mean single field inflation with a standard kinetic term, minimally coupled with gravity and no interactions with other particles.} reads $p = -X - V$). It is known that in the case of a minimal coupling (\emph{i.e.} for $f = 1$), gauge fields quickly decay during the Hubble expansion~\cite{Maleknejad:2011jr}. However, it has been shown that the presence of a non-minimal coupling between the inflaton and the gauge fields may sustain gauge field fluctuations during inflation (for a review of some of the possibilities see~\cite{Maleknejad:2012fw}). While in most of the existing literature it is common to assume a standard Lagrangian for the inflaton (\emph{i.e} $P = X +V $) and a non-minimal coupling $f(\phi, X)$ depending on $\phi$ only, generalizations of this framework have been discussed (see for example~\cite{Do:2011zz, Ohashi:2013pca, Do:2017qyd} for $P_{,X} \neq 1$ and~\cite{Holland:2017cza} for both $P_{,X} \neq 1$ and a derivative coupling\footnote{Motivations for this choice arise from string theory models of inflation where the parameterizations of $P(\phi, X)$ and $f(\phi, X)$ are specified from a Dirac-Born-Infeld (DBI) action~\cite{Dimopoulos:2011pe}.}).\\

In this work we discuss homogeneous configurations for both the inflaton and the gauge fields (meaning that both $A_{\mu}$ and $\phi$ depend on $t$ only) and quantum perturbations around these backgrounds. In particular, we will focus on two different cases:
\begin{enumerate}
	\item \label{cases:anisotropic} \textbf{Anisotropic ansatz}:  as discussed in~\cite{Watanabe:2009ct}, by imposing temporal gauge (\emph{i.e. $A_0 = 0$}) a homogeneous gauge field can always be expressed as:
	\begin{equation}
	\label{eq:anisotropicgauge}
	A_{\mu}(t) = (0,v_A(t),0,0) \; ,
	\end{equation}
	where, without loss of generality, we have chosen the gauge field to be aligned along the $x$ axis. In this case a consistent choice of  the metric is:
	\begin{equation}
	\label{eq:anisotropicmetric}
	 g_{\mu\nu} \textrm{d}x^\mu \textrm{d}x^\nu= - \textrm{d}t^2 + a^2(t) \left[ b(t)^{-4} \textrm{d}x^2 + b(t)^{2} \left( \textrm{d}y^2 + \textrm{d}z^2 \right) \right]\; ,
	\end{equation}
	where $a(t)$ and $b(t)$ are the isotropic and anisotropic scale factor respectively.
 	\item \label{cases:isotropic}\textbf{Isotropic ansatz}: 	similarly to the anisotropic case, $A_0  $ is set to zero by gauge choice, and we can proceed by considering the homogeneous background:
\begin{align}
	A^a_{i}(t)&= v_A(t)\delta^a_i\;,\label{eq:isotropicansatz}
\end{align}
which, together with the inflaton, evolves in a homogeneous and isotropic spacetime with metric:
\begin{align}
\label{eq:FLRW} g_{\mu\nu}\text{d}x^\mu \text{d}x^\nu&=-\text{d}t^2+a^2(t)\text{d}\vec{x}^2\;,
\end{align}
which is the usual FRLW background.
 \end{enumerate} 
In the remaining of this section we give a brief explanation~\footnote{More details on the derivation of the equations shown in this section are given in Appendix~\ref{appendix:general}.} of the general characteristics of the isotropic and anisotropic ansatz respectively.

\subsection{The isotropic ansatz}
\label{sec:isotropic} 
Two different mechanisms to realize an isotropic gauge field background are already present in the literature: 
 	\begin{itemize}
 		\item As discussed in~\cite{Golovnev:2008cf}, from the superposition of a large number $\mathcal{N}$ of homogeneous gauge fields (described by the ansatz of Eq.~\eqref{eq:anisotropicgauge}) with random directions, it is possible to obtain an approximately isotropic configuration (anisotropy scales like $1/\sqrt{\mathcal{N}}$). Moreover, phase-space analysis shows that the trajectories converge towards a universal isotropic attractor~\cite{Hervik:2011xm}.
 		\item As in the cases of Gauge-flation~\cite{Maleknejad:2011sq, Maleknejad:2011jw} (see also~\cite{Adshead:2012qe, SheikhJabbari:2012qf, Namba:2013kia}) or Chromo Natural Inflation (CNI)\footnote{For the analysis of  perturbations (which presents a quite reach and interesting phenomenology) see for example~\cite{Dimastrogiovanni:2012st, Dimastrogiovanni:2012ew, Adshead:2013qp, Adshead:2013nka, Domcke:2018rvv}.}~\cite{Adshead:2012kp}, by considering non-abelian $SU(2)$ gauge fields it is possible to perform a \emph{global} gauge transformation together with a coordinate transformation to align the three gauge fields along the three axes, obtaining an isotropic vacuum expectation value for the gauge fields\footnote{It is worth pointing out that both in CNI and Gauge-flation, the inflaton is assumed to be an axion which is coupled to the gauge fields through a $\phi F\tilde{F}$ term ($\tilde{F}$ being the dual field strength). However, the same procedure can analogously be performed with a scalar inflaton coupled to gauge fields as $f(\phi,X)FF$.}. In the limit of a sufficiently small gauge coupling, non-abelian interactions are suppressed and the three gauge fields effectively evolve as three abelian ($U(1)$)  fields which preserve isotropy.
 	\end{itemize}
 
 While different theoretical constructions are possible, in this work we focus our interest on phenomenology, and we do not discuss this point in detail. Such an analysis can be a interesting starting point for future works on topic. Assuming the isotropic ansatz specified by Eq.~\eqref{eq:isotropicansatz} and Eq.~\eqref{eq:FLRW}, $P = X + V$ and $f(X,\phi) = f(\phi)$ the equation of motion for inflaton and for the gauge fields read:
\begin{align}
	 \ddot{\phi}+3H\dot{\phi}+V_{,\phi} &=-\frac{1}{4}\left(-\frac{6 \dot{v}_A^2}{a^2}\right) f_{,\phi}  \;,\\
	\partial_{0}\left(a f \dot{v}_A\right) & =0\;, \label{eq:eom_gauge_isotropic}
\end{align}
and it is clear that Eq.~\eqref{eq:eom_gauge_isotropic} is solved by:
\begin{equation}
\label{eq:isotropic_gauge_sol}
\dot{v}_A = p_A a^{-1} f^{-1} \; , 
\end{equation}
where $p_A$ is a constant. This result can be substituted in the first equation and in particular, by defining the function $\tilde{f}\equiv fa^4$, we can express the only non-zero components of Einstein equations as:
\begin{align}
	\label{eq:friedman_isotropic}
	3H^2&= \frac{\dot{\phi}^2}{2} +V(\phi)+\frac{3p_A^2}{2\tilde{f}}\;,\\
	\label{eq:Raychauduri_isotropic}
	-2\dot{H}&=\dot{\phi}^2 +\frac{2p_A^2}{\tilde{f}}\;.
\end{align}
Once parameterizations for the scalar potential $V(\phi)$ and for $f(\phi)$ are specified, the system can be solved and the behavior of the solutions can be studied. In particular from the first slow-roll parameter $\epsilon_H$:
\begin{equation}
	\epsilon_{H} \equiv -\frac{\dot{H}}{H^2}  = \frac{\dot{\phi}^2}{2H^2}+\frac{p_{A}^2}{\tilde{f} H^2 }  = 3 \;  \frac{\dot{\phi}^2 + 2 p_A^2 / \tilde{f} }{ \dot{\phi}^2  +2 V(\phi) + 3 p_A^2 / \tilde{f} } \; ,
	\label{eq:epsh}
\end{equation} 
it is possible to understand whether inflation (\emph{i.e.} $\epsilon_{H} \ll 1$) is realized. As a matter of fact, inflation can only take place if the two conditions (recalling that $\tilde{f}>0$):
\begin{equation}
	\dot{\phi}^2  \ll V(\phi) \; , \qquad \qquad  p_A^2 / \tilde{f}  \ll V(\phi) \; , 
\end{equation}
are simultaneously satisfied. For later convenience it is also useful to introduce the two quantities:
\begin{equation}
\epsilon_{\phi}\equiv \frac{1}{2} \left(\frac{\dot{\phi}}{ H} \right)^2 \; , \qquad \qquad \rho_{A} = \frac{3 p_A^2}{2 \tilde{f}}  \; ,
\label{eq:epshmepsf}
\end{equation} 
where $\epsilon_{\phi}$ can be seen as the slow-roll parameter related to the inflaton field and $\rho_{A}$ is the gauge fields energy density.

\subsubsection{Isotropic power spectra}
\label{sec:isotropic_spectra}
Even if the gauge fields have a non-vanishing  energy density, at the level of classical background this would be indistinguishable from the single-field model\footnote{To the best of our knowledge, most (or all) of the typical gauge field induced backreaction can be mimicked by the introduction of particular features in the scalar potential.}. In order to extract information about the gauge field contribution, it is necessary to investigate its linear perturbation and its contribution to the scalar and tensor power spectra (higher order correlators, like the bispectrum, would give a further way to distinguish them). Similarly to~\cite{Ohashi:2013qba}, it is possible to express the power spectrum of curvature perturbations as a sum of two contributions coming from the inflaton field and from the (triad of) gauge fields:
\begin{equation}
\label{eq:isotropi_spec}
\left. \Delta_{s}^2 (k) \right|_{\frac{k}{ a H}  =1} =  \Delta_{s, 0}^2  \left[1 +  \frac{ 4 \rho_A }{ 3 H^2 }   \, \left(\frac{ \textrm{d} \ln f }{\textrm{d} \phi}  \right)^2  \,  N_k^2 \right]  \simeq  \Delta_{s, 0}^2 \left[1 +  \frac{ 32 \rho_A }{ 3 H^2 \epsilon_{\phi}  }   \,  N_k^2 \right] \;, 
	\end{equation}
where $\Delta_{s, \, (0)}^{2} \equiv H^2 / (8\pi^2 \epsilon_H)$ is the usual vacuum contribution to the scalar power spectrum and $N$ denotes the number of e-foldings defined as:
\begin{equation}
	N \equiv - \int_{a_{\textrm{f}}}^{a} \textrm{d} \ln a = - \int_{t_{\textrm{f}}}^t H \textrm{d} \ln t \; .
\end{equation}
In Eq.~\eqref{eq:isotropi_spec} $N_k$ denotes the number of e-foldings before the end of inflation at which the (gauge) modes with the wave number  $k$ left the horizon. Since in absence of multiple crossing\footnote{For the treatment of cases in which multiple crosses take place see~\cite{Ballesteros:2018wlw}. } this is uniquely determined by $k = aH$, from now on we drop the $k$ subscript (and also the $k/(aH) = 1$) and every time we present the scalar and tensor power spectra and derived quantities, we always intend them to be evaluated at horizon crossing. Notice that for $\rho_A = 0$ we consistently recover the usual expression for the scalar power spectrum. In order to compare with CMB observables it is customary to introduce the scalar spectral index defined as:
\begin{equation}
\label{eq:ns_isotropic}
	n_s \equiv 1 +\frac{\textrm{d} \ln \Delta_s^2}{\textrm{d} \ln k}  \simeq  1 - \left[2 \epsilon_{H} + \epsilon_{ 2} +  \frac{ 32 \rho_A N^2 }{ 3 H^2 \epsilon_{\phi}  }  \left( \frac{2}{N} -\frac{\epsilon_{\phi, N} }{\epsilon_{\phi} } - 2 \epsilon_{ H}  \right)  \right]  \; ,
\end{equation}
where ${}_{,N}$ denotes a derivative with respect to $N$ and where we define:
\begin{equation}
\label{eq:higher_slow_roll}
 \epsilon_{i + 1} = - \frac{\textrm{d} \ln |\epsilon_{i}| }{ \textrm{d} N } \; , \qquad \text{with} \;  \qquad    \epsilon_{0} \equiv \frac{1}{H}  \; , \qquad \text{and} \;  \qquad \epsilon_{1} \equiv \epsilon_{H} \; . 
\end{equation}
The first two terms appearing in the parenthesis in Eq.~\eqref{eq:ns_isotropic} are given by the usual vacuum contribution, the other terms are the modifications due to the presence of the inflaton-gauge field coupling. $n_s$ measures the tilt of the scalar power spectrum at a given scale and typically it is evaluated at CMB scales (corresponding to $N \simeq 60$) where the most recent observations by Planck~\cite{Akrami:2018odb} constrain its value to be $n_s \simeq 0.9649\pm 0.0042  $ at $68\%$ \rm{CL}. \\

 Analogously to the scalar case, the tensor power spectrum gets modified by the presence of the gauge fields, and it is equal to:
\begin{equation}
\Delta_{t}^2 =  \Delta_{t, \, (0)}^{2} \left(1+\frac{8 \rho_A}{3 H^2} \, N^2\right)\,,
\label{eq:isotensorps}
\end{equation}
where in analogy with the scalar case, we introduce $ \Delta_{t, \, (0)}^{2}  \equiv 2 H^2 / \pi^2 $ to denote the vacuum contribution to the tensor power spectrum.  In order to compare theoretical predictions with observations it is customary to introduce the tensor to scalar ratio as:
\begin{equation}
r\equiv  \frac{ \Delta_{t} ^2} { \Delta_{s}^2 }  \simeq 16 \; \epsilon_{H}  \left( 1 - \frac{ 32 \rho_A N^2 }{ 3 H^2 \epsilon_{\phi}  } \right)  \;,
\end{equation}
which shows that the tensor-to-scalar ratio is suppressed compared to the single-field slow roll inflation. When the gauge field does not contribute to the dynamics then $\rho_{A}=0$ and the tensor-to-scalar ratio reduces to $r = 16 \epsilon_H$.

\subsection{The anisotropic ansatz}
\label{sec:anisotropic}
Assuming the anisotropic ansatz specified by Eq.~\eqref{eq:anisotropicgauge} and Eq.~\eqref{eq:anisotropicmetric}, $P = X + V$ and $f(X,\phi) = f(\phi)$, the eom for the gauge fields reduces to (for details see  Appendix~\ref{appendix:general}):
\begin{align}
	\partial_0\left(ab^4f\dot{v}_A\right)=0\;, \label{eq:gauge_eq_anisotropic}
\end{align}
which is clearly solved by:
\begin{equation}
\dot{v}_A = p_A a^{-1} b^{-4} f^{-1} \; , 
\end{equation}
where $p_A$ is a constant. At this point, in analogy with the procedure carried out in the isotropic case, by defining $\tilde{f}=a^4b^4f$ the only non-zero components of Einstein equations read:
 \begin{align}
	3\left(H_a^2-H_b^2\right)&= \frac{\dot{\phi}^2}{2} +V(\phi)+\frac{ p^2_A  }{2 \tilde{f}}\;,\label{eq:anisofried1}\\
	3H_aH_b+\dot{H}_b&= \frac{p^2_A }{3 \tilde{f}}\;,\label{eq:anisofried2}\\
	3H_a^2+\dot{H}_a&=  V(\phi)+\frac{p^2_A }{6 \tilde{f}}\;, \label{eq:anisofried3}
\end{align}
where $H_a\equiv \dot{a}/a$ and $H_b\equiv\dot{b}/b$. Notice that for $p_A = 0$ (which corresponds to $\dot{v}_A=0$ \emph{i.e.} no gauge fields), we recover the usual case of the simplest realization of inflation. By combining Eq.~\eqref{eq:anisofried1} and Eq.~\eqref{eq:anisofried3} we can obtain the analogous of the $\epsilon_H$ parameter defined in the isotropic case:
\begin{equation}
\label{eq:epsilon_anisotropic}
 \epsilon_H \equiv -\frac{\dot{H}_a}{H_a^2} = 3 \left( \frac{H_b}{H_a} \right)^2 + \frac{p_A^2}{3 \tilde{f} H_a^2} + \frac{\dot{\phi}^2}{2 H_a^2}\; .
\end{equation}
Since for $\tilde{f} > 0$ this is a sum of positive terms, the condition to realize inflation is that all of the terms on the r.h.s. are much smaller than one. For later convenience we define:
\begin{equation}
	\label{eq:rho_A_def}
\epsilon_{\phi} \equiv \frac{1}{2} \left( \frac{\dot{\phi}}{H_a}\right)^2\; , \qquad \qquad	
\rho_A \equiv \frac{p_A^2}{2 \tilde{f}} \; ,
\end{equation}
which respectively correspond to the first slow-roll parameter in the isotropic case and  to gauge field energy density.

\subsubsection{Anisotropic power spectra}
\label{sec:anisotropic_power_spectra}
In the anisotropic case a privileged axis (here the $x-$axis) exists and the scalar and tensor power spectra are both characterized by a non-isotropic contribution~\cite{Dulaney:2010sq, Gumrukcuoglu:2010yc, Watanabe:2010fh, Soda:2012zm, Bartolo:2012sd} depending on the angle $\theta_{\bf{k},\bf{x}}$ between the wave vector $\bf{k}$ and the privileged direction $\bf{x}$. In particular, for the scalar power spectrum we have:
\begin{equation}
\Delta_{s}^2\left(\bf{k}\right) = \Delta_{s, (0)}^2  \,  \left[1+\frac{2 \rho_A}{H^2 }\left(\frac{ \textrm{d} \ln f }{\textrm{d} \phi} \right)^2 \,  N^2 \, \sin^2 \theta_{\bf{k}, \bf{x}} \right]  \,,
\label{eq:anisops}
\end{equation}
where $ \Delta_{s, (0)}^2 $ is the isotropic power spectrum. The extra contribution, compared to the vacuum case, is usually parameterized in terms of the anisotropic parameter $g_{*}$ introduced in~\cite{Ackerman:2007nb}  and defined as:
\begin{equation}
\label{eq:g_parameter}
|g_{*} |= \frac{16}{\epsilon_\phi}\frac{\rho_{A}}{H^2 } \,N^2\,,
\end{equation}
which make explicit that also a negligible contribution of the gauge field energy density generates anisotropic signatures in the spectra. $g_{*} $ encodes the level of anisotropy and it is well constrained to be $|g_{*}|\simeq 0.002\pm 0.016$ $(68\% \rm{CL})$~\cite{Ade:2015lrj, Ade:2015hxq, Kim:2013gka}. Even if observations strongly indicate a level of anisotropy compatible with zero, it is interesting to derive information about HJ variables from this constrain. This will be further developed in Sec.~\ref{sec:anisotropic_beta}. For completeness, we report the expression of the scalar spectral index, given by\footnote{After averaging $\sin^{2} \theta_{\bf{k_1,\bf{x}}}$ over all the angles}:
\begin{equation}
\label{eq:anisotropic_ns}
n_{s} \simeq 1-\left[2 \epsilon_{H} + \epsilon_{ 2} + \frac{2}{3} |g_{*}| \left(
	\frac{2}{N} - 2 \epsilon_{ H} - \frac{\epsilon_{\phi,N}}{\epsilon_{ \phi}}+ 2 \frac{\gamma_{,N}}{1 + \gamma}  \right)\right]  \;,
\end{equation}
where we introduced $\gamma \equiv H_b/H_a$ (more on this is said in Sec.~\ref{sec:anisotropic_beta}). Once again the first two terms in the bracket correspond to the usual vacuum contribution and the other terms are the gauge field-induced modifications.\\

Also the tensor power spectrum gets modified by a contribution that will be direction dependent~\cite{Watanabe:2010fh}. In particular, we find that the power spectrum of tensor perturbations, considering both polarizations, is:
\begin{equation}
 \Delta_{t }^2  \, =    \Delta_{t, (0)}^2  \,  \left(1 + \frac{ 4 \rho_{A}}{H^2} \,N^2 \sin^{2}\theta_{\bf{k},\bf{x}}\right)\;.
\label{eq:anisotensorps}
\end{equation}
Using Eq.~\eqref{eq:anisops} and Eq.~\eqref{eq:anisotensorps} we can compute the tensor-to-scalar ratio and relate it to the anisotropic parameter $g_*$:
\begin{equation}
r\equiv \frac{ \Delta_{t}^2  }{ \Delta_{s}^2  }=\, 16\epsilon_H \frac{1+ \frac{\epsilon_\phi}{4}  |g_* | \sin^{2}\theta_{\bf{k},\bf{x}} }{1+ |g_*| \sin^{2}\theta_{\bf{k},\bf{x}} }\,,
\label{tensortoscalar}
\end{equation}
which, averaging over all the angles, becomes:
\begin{equation}
r=\, 16\epsilon_H \frac{1+\frac{1}{6} \epsilon_\phi   |g_*| }{1+\frac{2}{3}  |g_*| } \simeq 16\epsilon_H \left(1- \frac{2}{3}  |g_*| \right) \,.
\end{equation}
As a consequence, the coupling between the inflaton and the gauge fields modifies the value of $r$. In particular, for larger values of $|g_*|$ at CMB scales, we expect smaller values of $r$. \\

Finally, it is worth pointing out that, as shown in~\cite{Bartolo:2012sd}, models with the interaction $\phi-F^2$, are characterized by a `consistency relation'  which relates the power spectrum to higher order correlation functions, like the bispectrum (\emph{i.e} the Fourier transform of the three-point correlation function). In particular, for the model analyzed in this paper, the anisotropy in
the power spectrum ($g_{*}$) is related to the bispectrum amplitude ($f_{NL}$)  (see for example~\cite{Bartolo:2012sd} for more details).

\section{$\beta$-function formalism: definition and applications}
 \label{sec:beta_function_general}
The $\beta$-function formalism is an alternative formulation of inflation which relies on the application of the HJ approach to cosmology~\cite{Salopek:1990jq}. Inspired by holographic arguments~\cite{Skenderis:2006jq, McFadden:2009fg, McFadden:2010na}, the $\beta$-function formalism was introduced in~\cite{Binetruy:2014zya} with the aim of defining a systematic (and theoretically well-motivated) classification of inflationary models. Interestingly, in a series of subsequent works~\cite{Pieroni:2015cma, Binetruy:2016hna, Cicciarella:2016dnv, Cicciarella:2017nls, Berera:2018tfc} it has been shown that the formalism provides a powerful tool to achieve a (semi-)analytical understanding of generalized models of inflation. In order to illustrate the method, we start this section with a brief review of its definition in the case of the simplest realization of inflation. The generalization to the isotropic and anisotropic ansatzes of Sec.~\ref{sec:model_definition} is then described in subsections~\ref{sec:isotropic_beta} and~\ref{sec:anisotropic_beta} respectively.\\

In the simplest realization of inflation the dynamics is completely specified by the equation of motion for the inflaton field:
\begin{equation}
\label{eq:klein_gordon_cold}
	\qquad \ddot{\phi} + 3H \phi +V_{,\phi} = 0 \;  , 
\end{equation}
and by (one of) the two non-zero component of Einstein equations:
\begin{eqnarray}
	\label{eq:einstein_cold}
	3H^2 &=& \rho \; , \\
  \label{eq:Raychauduri}
  -2 \dot{H} &=& p +\rho \; , 
\end{eqnarray}
where $p$ and $\rho$ are the pressure and energy density of the inflaton. Within the HJ formalism, under the reasonable assumptions that  a solution of this system exists and that the time evolution of $\phi$ as function of $t$ is piecewise monotonic, it is possible to invert to get $t(\phi)$ and use the field as a clock to describe the evolution of the system. At this point we introduce the so-called superpotential\footnote{ The formal analogy between the parameterization of the scalar potential in SUSY (for a review, see for example~\cite{Binetruy:2006ad}) and in the formalism~\cite{Binetruy:2014zya}, justifies the name.} $W(\phi)\equiv-2H(\phi)$ and using Raychauduri's equation~\eqref{eq:Raychauduri} we directly find:
\begin{align}
	\dot{\phi}=W_{,\phi}\;, \label{eq:phidotphi_cold}
\end{align}
implying that it is possible to express $\dot{\phi}$ (and therefore all physical quantities) as a function of $\phi$ only. It is worth stressing that this is a crucial step for the definition of the formalism and, as a matter of fact, this can be non-trivial for generic $P_{,X} \neq 1$ (and/or $f_{,X} \neq 0$). Indeed, this motivates our choice of avoiding unnecessary complications by assuming $P_{,X} = 1$ and $f_{,X} = 0$. However, we point out that it was shown in~\cite{Binetruy:2016hna} that for single field models of inflation the formalism can successfully incorporate models with $P_{,X} \neq 1$. By following a similar procedure, a further generalization of the formalism which includes the cases neglected in the present work could be defined. \\

By following a formal analogy with the definition of the RGE describing the evolution of the renormalized coupling constant (whose role here is played by the inflaton) in terms of the renormalization scale (here the scale factor $a$) we introduce the cosmological $\beta$-function as:
\begin{equation}
	 \label{eq:beta_function_definition}
	 \beta(\phi) \equiv \frac{\textrm{d} \phi}{\textrm{d} \ln a} \; .
\end{equation}
From this definition and using Eq.~\eqref{eq:phidotphi_cold}, Eq.~\eqref{eq:einstein_cold} and Eq.~\eqref{eq:Raychauduri} it is easy to show that:
\begin{equation}
	 \label{eq:eq_of_state_cold}
	 \frac{p+\rho}{\rho} = \frac{4}{3} \, \frac{W_{,\phi}^2 }{
           W^2} = \frac{\beta^2 (\phi)}{3} \; .
\end{equation}
This expression for the equation of state shows that a phase of nearly exponential expansion of the Universe (such as inflation) corresponds to the neighborhood of a zero of $\beta(\phi)$. In particular, a de Sitter phase of expansion corresponds to a fixed point of the associated RG flow (\emph{i.e.} a point where $\beta(\phi) = 0$). The leading order characterization of $\beta(\phi)$ around the fixed point uniquely defines a class of models of inflation. Higher order contributions to the $\beta$-function, which are not playing any role in the neighborhood of fixed point, may become important along the RG flow and break the universality at different scales. This clarifies the power of this formalism since all the information on the inflationary phase is  enclosed in the parameterization of $\beta(\phi)$.\\

 Clearly, the presence of a gauge field contribution requires some modifications in the definition of the formalism (for example Eq.~\eqref{eq:phidotphi_cold} have different solutions). However, as shown in the following sections, until the evolution of the inflaton is piecewise-monotonic in time, all the relevant quantities can still be expressed as functions of the field only. As an instructive example, we first consider the simpler case of the isotropic configuration~\eqref{eq:isotropicansatz}. After that the formalism for the system inflaton-gauge fields is established we finally proceed to the extension to the anisotropic ansatz~\eqref{eq:anisotropicgauge}-\eqref{eq:anisotropicmetric}. 

\subsection{$\beta$-function formalism for the isotropic ansatz}
\label{sec:isotropic_beta}
Following an analogous of the procedure used for single-field inflation, we introduce the superpotential $W(\phi) \equiv -2 H(\phi)$ and we express Eq.~\eqref{eq:Raychauduri_isotropic} as: 
\begin{equation}
	\dot{W}(\phi)=W_{,\phi} \dot{\phi} = \dot{\phi}^2 +\frac{2p_A^2}{\tilde{f}} = \dot{\phi}^2 +\frac{4 \rho_A}{3} \;. \label{eq:phidotiso}
\end{equation}
We can now solve this equation for $\dot{\phi}$ to get:
\begin{align}
	\label{eq:dot_phi_isotropic}
	\dot{\phi}&=\frac{1}{2}W_{,\phi}\left(1+\sqrt{1-\frac{16 \rho_A}{3 W_{,\phi}^2}}\right)\,,
\end{align}
which, by definition, is a function of the scalar field only. Notice that when the second term in the square root vanishes we recover the usual $\dot{\phi} = W_{,\phi}$. By defining the $\beta$-function as in Eq.~\eqref{eq:beta_function_definition} and using Eq.~\eqref{eq:dot_phi_isotropic} we get:
\begin{align}
	\label{eq:beta_isotropic}
	\beta(\phi) &= -\frac{W_{,\phi}}{W}\left(1+\sqrt{1-\frac{16 \rho_A}{ 3 W_{,\phi}^2}}\right) = \frac{\beta_0 }{2 } \left(1+\sqrt{1-\frac{64 \rho_A}{3 \beta_0^2 W^2 }}\right)\; ,
\end{align}
where, in analogy with the procedure of~\cite{Berera:2018tfc}, we have introduced:
\begin{equation}
	\label{eq:beta_0_superpot_iso}
	\beta_{0}\equiv -2 \frac{W_{,\phi}}{W} \; , \qquad \text{implying} \qquad W =W_f\exp\left\{-\frac{1}{2}\int_{\phi_f}^\phi\text{d}\phi'\beta_0(\phi')\right\} \; . 
\end{equation}
Notice that for $\rho_A= 0$ Eq.~\eqref{eq:beta_isotropic} gives the usual $\beta = \beta_0 = -2 W_{,\phi} /W$. At this point it is interesting to stress that Eq.~\eqref{eq:dot_phi_isotropic} (and as a consequence Eq.~\eqref{eq:beta_isotropic}) only makes sense if the condition:
\begin{equation}
\label{eq:condition_isotropic}
\beta_{0}^2 \equiv 4\frac{W_{,\phi}^2}{W^2} \geq \frac{64 \rho_A}{3 W^2} 	\; , 
\end{equation}
is satisfied. Such a constraint, which is not present in the standard formalism, naturally emerges in this framework. Indeed, by solving Eq.~\eqref{eq:Raychauduri_isotropic} for $\dot{\phi}$, the formalism automatically contains information on the dynamics of the system. Since specifying an inflationary model consists in fixing the parameterisation of  $\beta_0(\phi)$ (or of $\beta(\phi)$) and $f(\phi)$ plus some initial conditions (\emph{i.e.} $p_A$ and $W_\textrm{f}$), Eq.~\eqref{eq:condition_isotropic} can be seen as a straightforward sanity check for the model. This clearly shows that the formalism provides a powerful tool for model building.\\

In order to have a better understanding of the link between the present case and simplest realization of inflation, it is useful to notice that, if the condition in Eq.~\eqref{eq:condition_isotropic} is satisfied, we have:
 \begin{equation}
 |\beta_0|/2 \leq |\beta| \leq |\beta_0| \; .
 \end{equation}
At this point, by computing the expression of the first slow-roll parameter:
 \begin{equation}
 \epsilon_H \equiv -  \frac{\dot{H}}{H^2} =  \frac{1}{2} \beta \beta_0  = \frac{1}{4}\beta_{0}^2\left(1+\sqrt{1-\frac{64 \rho^2_A}{3\beta_0^2W^2}}\right)\; , \label{eq:epsilon_beta}
 \end{equation}
it becomes manifest that, until Eq.~\eqref{eq:condition_isotropic} is satisfied, inflation can only be realized for $\beta_0 \ll 1$ (and thus $\beta \ll 1$). Notice that if $\rho_A = 0$, inflation ends for $\beta_0 = \sqrt{2}$, on the other hand, if the square root is zero, inflation ends for $\beta_0 =2$ implying that, similarly to~\cite{Barnaby:2011vw, Barnaby:2011qe, Domcke:2016bkh}, the gauge field production is effectively modifying the last part of the evolution.  Moreover, it is also interesting to consider the expression of the scalar potential:
\begin{align}
\label{eq:isotrpopic_pot}
V(\phi)&=\frac{3}{4}W^2(\phi)\left[ 1-\frac{1}{6}\beta^2(\phi) -\frac{2 p_A^2}{ \tilde{f} W^2 } \right] \; .
\end{align}
As Eq.~\eqref{eq:condition_isotropic} implies that the last term in the bracket is (at least) of order $\beta_0^2$ (or higher), and $\beta$ is of the same order of $\beta_0 $ (see Eq.~\eqref{eq:beta_isotropic}), the parameterization of $\beta_0$ (which sets $W(\phi)$ through Eq.~\eqref{eq:beta_0_superpot_iso}) directly specifies $V(\phi)$. As a consequence, even in the presence of gauge fields, $\beta_0$  can still be used to specify (classes of) single-field models of inflation. \\

Similarly to the case of~\cite{Berera:2018tfc}, the formalism offers a natural representation of the system in a two-dimensional plot capturing the deviation from the simplest realization of inflation for different parameterization of the coupling constant. This becomes clear if, using the last equality in Eq.~\eqref{eq:phidotiso},  we recast Eq.~\eqref{eq:epsilon_beta} as:
\begin{equation}
\epsilon_{H} \equiv - \frac{\dot{H}}{H^2 } = \frac{ \beta^2 }{2} +  \frac{8 \rho_A}{3 W^2}  \; .
\end{equation}
Concretely, setting the square root of the two terms on the r.h.s. of this equation as the axis of the plot, we represent inflation as a curve going from the dS fixed point (corresponding to $(0,0)$) to a point of the unit-circle (corresponding to the end of inflation) in the $(+,+)$ quadrant of the plane. For example, if the contribution of the gauge field is zero, we expect an horizontal trajectory and as soon as the contribution of the gauge field start to be non-negligible we expect curves departing from this horizontal line. The shape of the different trajectories will be mostly defined by the coupling function $\tilde{f}$. For more details on such analyses we refer the reader to~\cite{Berera:2018tfc}. \\

While in this section we have mainly focused on the evolution of the background, the formalism also provides a powerful tool to characterize perturbations. In particular, as we will see for the anisotropic case, the equations provided in Sec.~\eqref{sec:anisotropic_power_spectra} will be expressed in terms of the typical quantities of the formalism. 

\subsection{$\beta$-function formalism for the anisotropic ansatz}
\label{sec:anisotropic_beta}
In this section we discuss the extension of the $\beta$-function formalism to include the anisotropic ansatz introduced in the bullet point~\eqref{cases:anisotropic} of Sec.~\ref{sec:model_definition}. This is achieved by following a procedure similar to the one carried out in the previous section. However, it is interesting to point out that a crucial difference with respect to such a case is the presence of a second scale factor (the function $b(t)$), which requires the definition of a second superpotential. For better convenience, we connect the evolution of the two superpotentials via a new function $\gamma$, which measures the deviation between the isotropic and anisotropic components. Some comments on the interpretation of such function in the context of holography are given. \\

Considering the metric~\eqref{eq:anisotropicmetric}, in this case there are two different scale factors so we introduce two superpotential:
\begin{equation}
\label{eq:superpo_anisot}
 W_a(\phi)\equiv -2H_a(\phi) \; , \qquad \qquad W_b(\phi)\equiv-2H_b(\phi) \; , 
\end{equation}
 \emph{i.e.}, in addition to the usual superpotential (here called $W_a$), we introduce the `anisotropic' superpotential $W_b$. The next step towards the formulation of the problem in terms of the HJ formalism, is to express $\dot{\phi}$ as a function of $\phi$ only. Starting from the system in Eqs.~\eqref{eq:anisofried1}-\eqref{eq:anisofried3}, and using the definitions of Eq.~\eqref{eq:superpo_anisot}, with some algebra we get:
\begin{align}
	\dot{\phi}^2 -\left( {W}_{a,\phi} + {W}_{b,\phi}\right) \dot{\phi} +\frac{3}{2} W_a W_b +\frac{3}{2} W_b^2 & = 0\;.
\end{align}
This is a polynomial equation for $\dot{\phi}$ which is solved by:
\begin{align}
	\dot{\phi}&=\frac{1}{2}\left(W_{a,\phi}+W_{b,\phi}\right)\left[1 + \sqrt{1-6\frac{W_b}{W_a}\left(1+\frac{W_b}{W_a}\right)\left(\frac{W_{a,\phi}}{W_a}+\frac{W_b}{W_a}\frac{W_{b,\phi}}{W_b}\right)^{-2}}\right]\; , \label{eq:phidotWaWb}
\end{align}
where we have picked the plus solution in order to ensure consistency with the standard single-field case (i.e. $W_b = 0$ and $W_{a,\phi} = \dot{\phi}$). Also in this case we can state that $\dot{\phi}$ can be expressed as a function of $\phi$ only, which ensures that the HJ formalism can be applied even in the case where there is an anisotropic contribution to the dynamics of the system. \\

When anisotropic model of inflation, like the one in Eq.~\eqref{eq:general_action}, are considered, they are uniquely specified by a choice for the scalar potential $V$ and for the coupling function $f$, that in principle is function of $\phi$. The evolution of the system is completely fixed as soon as these two functions are specified. Since the behaviour of the coupling function $f$ defines the fate of the anisotropic expansion, we can think in the $\beta$-function framework, that we trade the potential for the $\beta$-function introduced in Eq.~\eqref{eq:beta_function_definition} and the coupling function for the new $\gamma$-function:
\begin{align}
 \gamma(\phi)\equiv \frac{\textrm{d} \ln b}{\textrm{d} \ln a} &= \frac{W_b}{W_a} \;,
\end{align}
which grasps the grows of the anisotropies. To the best of our knowledge, this is the first time that such a function, whose introduction is a minimum requirement to study non-isotropic fields within the HJ formalism, is introduced for studying the evolution of anisotropies during inflation. It is however fair to point out that such a procedure is already been developed in the exploration of gauge/gravity correspondence~\cite{Maldacena:1997re} (see for example~\cite{Gursoy:2018umf}). In such a framework an analogous\footnote{In this context the action of Eq.~\eqref{eq:general_action} (with $f= \exp(- \lambda \phi)$ and $P = X +V$) is typically referred to as a Einstein-Maxwell-Dilaton theory. Notice however that in the corresponding analyses this action is not used to consider anisotropic cosmologies but rather black hole solutions.} of the action in Eq.~\eqref{eq:general_action} is typically used for the holographic modeling of strongly interacting condensed matter systems~\cite{Taylor:2008tg, Hartnoll:2009sz, Hartnoll:2009ns, Charmousis:2010zz, Gouteraux:2011ce}. In terms of the gauge/gravity correspondence the models described by the action in Eq.~\eqref{eq:general_action} can thus be seen as the dual of a theory with Lifshitz scaling in the space components~\cite{Gouteraux:2011qh, Kiritsis:2012ma, Gouteraux:2012yr}. As a consequence, the $\beta$-function captures the evolution of the isotropic scale factor which in the context of holography is connected to the renormalization scale in the dual theory. Conversely, the $\gamma$-function is related to the growth of anisotropies which traces back to the introduction of a non-zero temperature and of a finite charge density in the dual QFT (and more in general to the breaking of Lorentz symmetry into Lifshitz~\cite{Korovin:2013bua, Korovin:2013nha}). \\

Following the procedure of the previous section (in particular Eq.~\eqref{eq:beta_0_superpot_iso}) we introduce the function $\beta_0$: 
\begin{equation}
	\label{eq:beta_0_superpot_aniso}
	\beta_{0}\equiv -2 \frac{W_{a,\phi}}{W_a} \; .
\end{equation}
This choice allows to directly compute $W_a$ and moreover it will become evident later that this function allows to directly generalize the classes of universality introduced for single field inflation to anisotropic scenario. Let us show that it is sufficient to specify $\beta_0$ and $\gamma$ to completely fix the model. Using Eq.~\eqref{eq:phidotWaWb} (and also Eq.~\eqref{eq:gamma_phi}) it is possible to show that $\beta$ can be re-expressed as:
\begin{equation}
\beta(\phi) \equiv - 2 \frac{\dot{\phi} }{W_a} =\left[ \frac{\beta_0 }{2} \left( 1 + \gamma \right) - \gamma_{,\phi}  \right] \left[ 1 + \sqrt{\Delta} \right]  \;,
\end{equation}
where we have introduced the definition:
\begin{equation}
\Delta \equiv 1 - 6 \gamma (1 +\gamma) \left[ -\frac{\beta_0}{2} \left( 1 + \gamma \right)  + \gamma_{,\phi} \right]^{-2} \;.
\end{equation}
Notice that $\Delta $ corresponds to the square root appearing in Eq.~\eqref{eq:phidotWaWb}. At this point it becomes easy to notice that the condition for inflation in term of $\beta_0$ and $\beta$ simply reads:
\begin{align}
	\epsilon_H &= -\frac{\dot{H}_a}{H^2_a} = \frac{1}{2}\beta_0\beta \ll 1\;,
	\label{epsilonbeta}
\end{align}
which shows explicitly that the different parametrisation of $\beta_0$ (and therefore the corresponding classes of universality found in the standard case) can be directly studied in the anisotropic scenario. \\

We are now in a position to express the conditions for the anisotropic system to describe an inflating universe in terms of our formalism.  From Eq.~\eqref{eq:epsilon_anisotropic} this automatically translates in the following conditions (for details see Appendix~\ref{appendix:some_more_eqs}):
\begin{align}
\gamma &\ll 1 \;  , &&  \gamma \left[ 1 - \frac{ \beta }{3} \left( \frac{\beta_0}{2} - \frac{\textrm{d} \ln \gamma }{\textrm{d} \phi } \right) \right] \ll 1 \; , && \beta^2 \ll 1 \; .
\label{eq:anisobetaconditions1}
\end{align}
Moreover, another condition which must be satisfied for our model to be consistent is that the square root in Eq.~\eqref{eq:phidotWaWb} is well defined, meaning that:
\begin{align}
\Delta & =1 -  \frac{6 \gamma}{1 + \gamma} \left[ \frac{\beta_0}{2} - \frac{\textrm{d} \ln \left( 1 + \gamma \right) }{ \textrm{d}\phi}  \right]^{-2} > 0\;.\label{eq:anisobetaconditions2}
\end{align}
Since $W_a $ and $W_b$ have the same sign, $\gamma $ is positive, the second term of the previous equation is negative and thus $\Delta $  is always smaller than one. Interestingly, there are three possibilities to get sensible inflationary solutions:
\begin{enumerate}
\item $ \beta_0 \ll \gamma_{,\phi}$ so that  the condition Eq.~\eqref{eq:anisobetaconditions2} reduces to $\gamma_{,\phi}^{\, 2} \gtrsim 6 \gamma$.
\item $\beta_0 \sim \gamma_{,\phi}$ which again requires $\gamma_{,\phi}^{\, 2} \gtrsim 6 \gamma$, and thus it can be seen as a particularly case of the previous condition. 
\item  $ \beta_0 \gg \gamma_{,\phi}$ so that  the condition Eq.~\eqref{eq:anisobetaconditions2} reduces to  $\beta_0^2  \gtrsim 24 \gamma$ .
\end{enumerate}

The first of these possibilities corresponds to the case in which $\gamma$ (and correspondingly the anisotropies) grows fast along the inflationary trajectory. As $\gamma$ must be smaller than one throughout the evolution, this implies that the initial amount of anisotropies must be strongly suppressed.  Moreover, some problems might arise from the second condition in Eq.~\eqref{eq:anisobetaconditions1}. This requires further checks on the allowed parameterizations for $\beta$. Similar problems may potentially arise for the second possibility as well. Conversely, the third possibility correspond to small anisotropies (at least second order in $\beta_0$)  in the deep inflationary phase (near the dS fixed point) which however may grow sufficiently fast during the last part of the evolution. Since solutions of the first two kinds may be more involved than the latter, in the following we only restrict our analysis to this case.\\

As discussed in Sec.~\ref{sec:anisotropic_power_spectra}, the presence of an anisotropic field induces statistical anisotropy in scalar and tensor power spectra. The information from CMB and LSS observations about anisotropy can then be used to set constrains on the HJ variables. Using expressions for the gauge field energy density and for the slow-roll parameter (see Eq.~\eqref{gaugeenergybeta} and  Eq.~\eqref{epsilonbeta} respectively), the scalar power spectrum reads:
	\begin{equation}
	\Delta_{s}^2\left(\bf{k}\right) \simeq \frac{W_a^2}{16 \pi^2 \beta \beta_0}  \,  \left(1+ |g_*|\, \sin^2 \theta_{\bf{k}, \bf{x}} \right)  \,,
	\end{equation}
where the parameter $g_{*}$ of Eq.~\eqref{eq:g_parameter} in  term of the new variables $\beta$ and $\gamma$ reads:
\begin{equation}
|g_{*} |= \frac{144 \gamma (1+\gamma)^2}{\beta^2} \left[1- \frac{\beta}{3}\left(\frac{\beta_0}{2} - \frac{\textrm{d}\, \ln \gamma}{\textrm{d}\, \phi}\right)\right]  N^2 \;.
\end{equation}

Following a similar procedure all the other quantities given in Sec.~\ref{sec:anisotropic_power_spectra} can be expressed in terms of the formalism. Some of these equations are reported in Appendix~\ref{appendix:general} (see in particular Sec.~\ref{appendix:some_more_eqs}). A similar treatment can also be applied to higher order correlation functions (\emph{i.e.} non-Gaussianity), whose discussion is given in~\cite{Bartolo:2012sd}. The information brought by non-Gaussianity could be used to get a further characterization of the HJ variables. Since these topics are beyond the scope of the present work, we conclude this section by considering an explicit example to show the effectiveness of the method.

\subsubsection{An explicit example - Anisotropic chaotic inflation }
\label{sec:anisotropic_chaotic}
 As explained in the previous sections, in the context of the $\beta$-function formalism a model is completely specified by fixing a choice for $\beta_0$ (or alternatively $\beta$) and for $\gamma$. In order to generalize the classes of universality defined in the standard case, without any scalar-gauge interaction, we proceed by specifying $\beta_0$. For this example we choose a function of the form:
\begin{align}
	\beta_0(\phi) &= -\frac{\alpha}{\phi}\;, \label{eq:beta0exp}
\end{align}
which corresponds to chaotic models of inflation with scalar potential $V (\phi) \simeq V_0 \, \phi^\alpha$. The possible choices for the  $\gamma$-function are then constrained by the conditions in Eq.~\eqref{eq:anisobetaconditions1} and in  Eq.~\eqref{eq:anisobetaconditions2}. A possible parametrization which respects this constraints is:
\begin{align}
	\gamma(\phi) & = C \left|\beta_0(\phi) \right|^\delta\;,\label{eq:gammaexp}
\end{align}
where $C > 0$ and $\delta > 2$ are two positive constants. In particular, this choice ensures $\Delta > 0$ during inflation. Notice that once the constants $\alpha$ in $\beta_0$ and $\delta$ in $\gamma$ have been fixed (\emph{i.e.} a class of models is specified), a variation in the third constant $C$ would directly correspond to a variation of the anisotropic parameter $g_{*}$.\\

\begin{figure}[t!]
    {\includegraphics[width=.475\columnwidth]{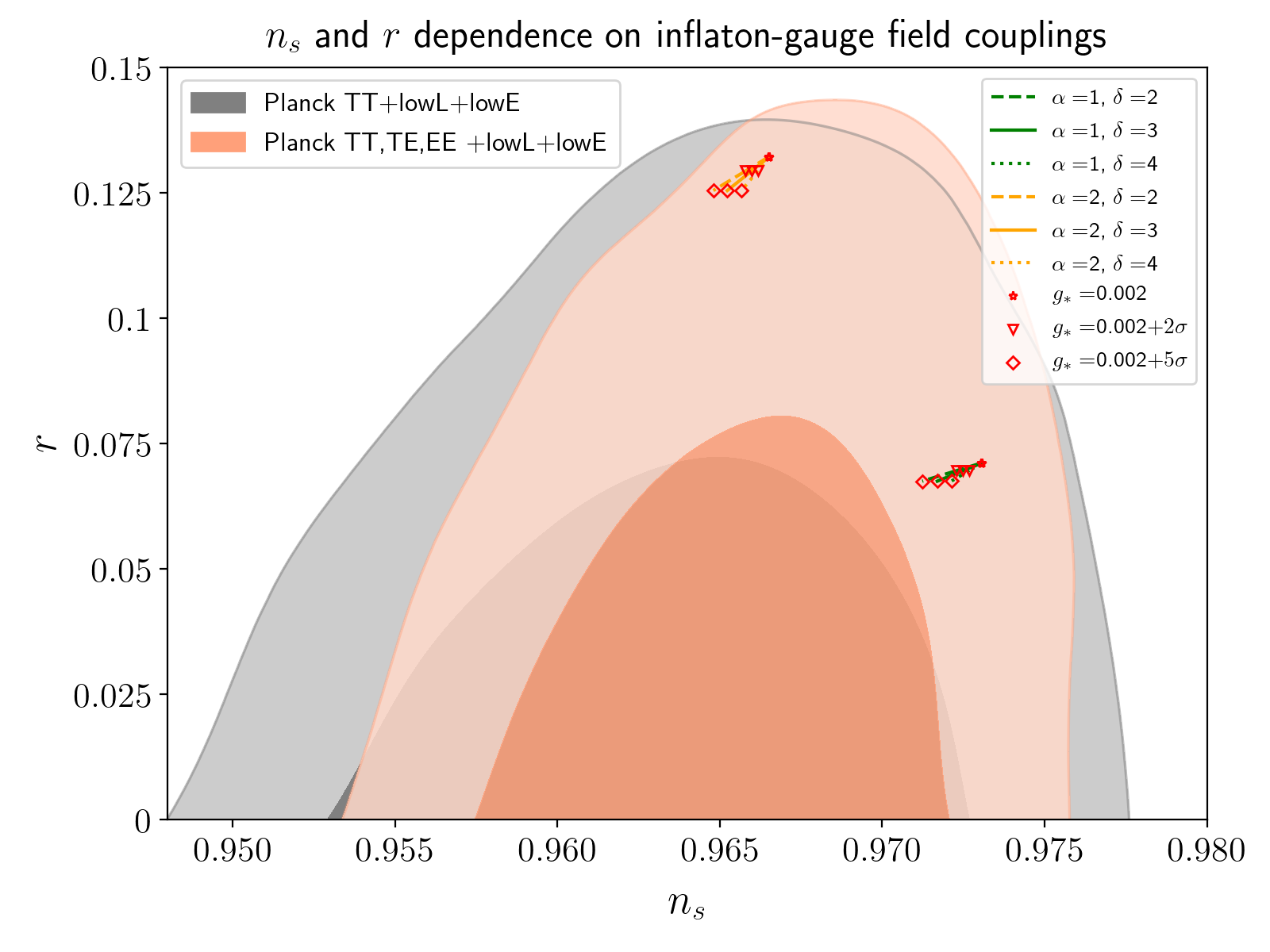}}
    \quad
        {\includegraphics[width=.475\columnwidth]{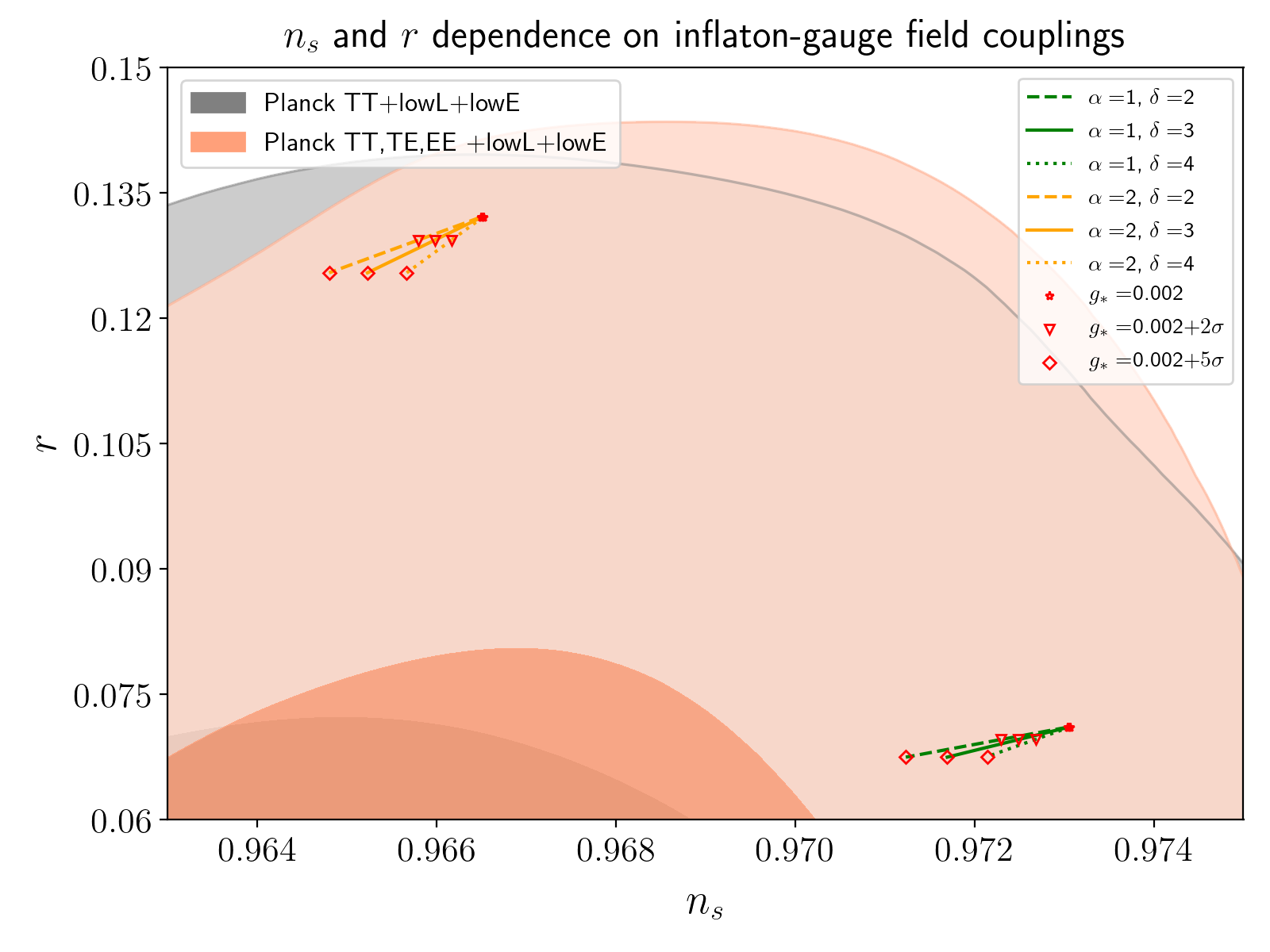}}
\caption{\label{fig:ns_r_evolution} Predictions for $n_s$ and $r $ in  linear scale for a set of models of the chaotic class with $\beta_0$ as in Eq.~\eqref{eq:beta0exp}, $\alpha = [1, 2 ]$ and a $\gamma$-function as in Eq.~\eqref{eq:gammaexp} with $\delta = [2,3,4]$ and the constant $C$ chosen to give and maximal value of  $|g_*|= 0.002+5\sigma$, with $\sigma=0.016$. Experimental constrains from the Planck satellite~\cite{Akrami:2018odb} are plotted in the background. The right panel is a zoom of the relevant region in the left panel. }
\end{figure}

Fig.~\ref{fig:ns_r_evolution} shows the variation in the predictions for the scalar spectral index $n_s$ and the tensor to scalar ratio $r$ in correspondence to the variation of $C$ for different choices of parameters $\alpha$ and $\delta$. The corresponding values of $n_s$ and $r$ are compared with the one and two sigma regions obtained by the Planck data~\cite{Akrami:2018odb}. The constant $C$ in Eq.~\eqref{eq:gammaexp} has been chosen such that the $|g_*|$ is never larger than five standard deviations from its best-fit value $0.002$. The predictions at $|g_*|=[0.002, 0.002+2\sigma,  0.002+5\sigma]$ with $\sigma=0.016$~\cite{Kim:2013gka} have been explicitly indicated on the plots. We observe that the interaction between the scalar and gauge fields produce a very small variation, in particular a decrease, in both the scalar spectral index and the tensor to scalar ratio.

\section{Conclusions}
\label{sec:conclusions}
 In this paper we have applied the $\beta$-formalism of~\cite{Binetruy:2016hna} to inflationary scenarios where the inflaton is coupled to homogeneous gauge fields~\cite{Watanabe:2009ct}. Such a formalism relies on a formal resemblance between the equations describing the evolution of the homogeneous inflaton in cosmology and RGE in QFT. This method turns out to be extremely efficient to define set of universality classes for inflationary models based on their scaling properties of the system. In concrete, we have considered the presence of $U(1)$ gauge field(s) and we studied the corresponding modifications in the definition of the formalism. We have considered both the case where the interplay of (three) gauge fields can generate isotropic configurations and, more interestingly, the case where the gauge field expands in a Bianchi type I metric leaving distinctive features in the spectrum of both scalar and tensor perturbations.  In particular, for the anisotropic ansatz described in Sec.~\ref{sec:isotropic_beta}, the presence of a preferred direction (and thus of a second scale factor, besides the isotropic one), imprinted by the spin-1 nature of the field, requires the introduction of a second superpotential. The evolution of the two superpotentials are related by a new function $\gamma$ which measures the growth of anisotropies. This is the first time that such a function is introduced for studying the evolution of anisotropies during inflation. It is worth mentioning that an extensive treatment of models similar to the one considered in this paper can be found in the context of holography applied to condense matter physics (see for example~\cite{Taylor:2008tg, Hartnoll:2009sz, Hartnoll:2009ns, Charmousis:2010zz, Gouteraux:2011ce}). \\

We have derived the conditions for both the isotropic and anisotropic configurations to describe an inflating universe, and we have translated such conditions in term of the $\beta$-function formalism. Moreover, we have shown that dynamical constraints, which provide a useful guideline for model building, naturally arise within the formalism. The formalism easily allows to study the influence of the gauge field on the inflaton dynamics. In particular, it can be used to identify the regions in the space of parameters where the coupling function $f$ can change the inflationary trajectories and consequently the (CMB) observables. In fact, models which are characterized by non-isotropic evolution, leave particular signature in the CMB, like a well know quadrupolar modulation in the spectrum of temperature fluctuations. CMB and LSS experiments gave stringent constraints on such a level of anisotropy, which we have used in this paper to derive information on the HJ variables.	\\

In order to study the role of the gauge fields during inflation, in this work we assumed the inflaton sector to be as minimal as possible \emph{i.e.} we assumed a single scalar field, a standard kinetic term and a minimal coupling to gravity. In building a fully realistic model for inflation one, or possibly all, of these assumptions may be relaxed. While some generalizations of the $\beta$-function formalism to more realistic landscapes have already been discussed\footnote{See~\cite{Pieroni:2015cma} for non-minimal couplings to gravity and~\cite{Binetruy:2016hna} for non-standard kinetic terms.}, a vast amount of work still needs to be carried out in this direction. For example, the exploration of multi-field models of inflation\footnote{The discussions of~\cite{Bourdier:2013axa, Garriga:2014fda, Garriga:2015tea} could provide useful guidelines for this purpose.} or more general kinetic couplings between the inflaton and gravity\footnote{Like in Horndeski's theories~\cite{Horndeski:1974wa} such as `New Higgs' models~\cite{Germani:2010gm, Germani:2010ux}.}. Another interesting possibility would be to apply the HJ formalism to scenarios where different background symmetry pattern are considered during inflation, like Solid Inflation and its extensions~\cite{Endlich:2012pz, Bartolo:2013msa, Bartolo:2014xfa, Bartolo:2015qvr, Ricciardone:2016lym}. Even if, in this class of models, the background fields are space-dependent, it is still possible to find combinations of quantities that are time dependent on the background, which is a necessary condition for the formalism to be applied. This could be an interesting topic for future investigations.

\vspace{1cm}
\subsubsection*{Acknowledgements}

We thank E.~Kiritsis and  L. Silva Pimenta for discussion and valuable comments.  J.M. is supported by Principals Career Development Scholarship and Edinburgh Global Research Scholarship. J.M. would like to thank the Instituto de Fisica Teorica (IFT UAM-CSIC) in Madrid for its support and hospitality. J.M. and M.P. acknowledge the support of the Spanish MINECOs ``Centro de Excelencia Severo Ocho'' Programme under grant SEV-2016-059. This project has received funding from the European Unions Horizon 2020 research and innovation programme under the Marie Sk\l{}odowska-Curie grant agreement No 713366.

\newpage 

\appendix

\section{General equations}
\label{appendix:general}
In this appendix we provide some of the details of the computations presented in Sec.~\ref{sec:model_definition} and in Sec.~\ref{sec:beta_function_general}. Let us start by computing the equations of motion for the inflaton field and for the gauge field described by the action of Eq.~\eqref{eq:general_action}:
\begin{align}
\label{eq:eom_inflaton_general}
	\frac{1}{\sqrt{|g|}}\partial_\mu\left[\sqrt{|g|}\left(P_{,X}+\frac{1}{4}f_{,X}F^2 \right)\partial^\mu\phi \right] - P_{,\phi} - \frac{1}{4}f_{,\phi}F^2 &= 0 \;,\\
	\label{eq:eom_gauge_general}
	\partial_{\mu} \left[ \sqrt{|g|} f F^{a,\mu\nu} \right] & =0\;.
\end{align}
Moreover, the stress-energy tensor $T_{\mu\nu}$ associated with the system reads:
\begin{equation}
\label{eq:stress_energy_general}
T_{\mu\nu} = \partial_\mu\phi\partial_\nu\phi \left(P_{,X}+\frac{1}{4}f_{,X}F^2 \right) +f g^{\rho\sigma}F^a_{\mu\rho}F^a_{\nu\sigma}-g_{\mu\nu}\left(P+\frac{1}{4}fF^2 \right)   \;.
\end{equation}
Assuming both the inflaton field and the metric to be homogeneous the e.o.m. for the inflaton reads:
\begin{equation}
	 \frac{\textrm{d} }{\textrm{d} t} \left[  \sqrt{|g|} P_{,X} \dot{\phi} + \frac{1}{4} f_{,X}  \dot{\phi} F^2 \right]  + \sqrt{|g|} P_{, \phi } + \sqrt{|g|} \frac{1}{4} f_{,\phi} F^2 = 0  \; .
\end{equation}
Reducing to the case of a standard Lagrangian for the inflation field (\emph{i.e.} $P=X+V(\phi)$) the system further simplifies to: 
\begin{align}
\label{eq:eom_inflaton_standard_P}
	\ddot{\phi}+\frac{1}{\sqrt{|g|}}\frac{\partial\sqrt{|g|}}{\partial t}\dot{\phi}+V_{,\phi} + \frac{1}{4}(f_{,X}\ddot{\phi}+f_{,\phi})F^2 +\frac{1}{\sqrt{|g|}}\frac{\partial}{\partial t}\left(\frac{1}{4}\sqrt{|g|}f_{,X}F^2\right)\dot{\phi} & = 0 \;,\\
 \partial_\mu\phi\partial_\nu\phi \left(1+\frac{1}{4}f_{,X}F^2\right)-g_{\mu\nu}\left(X+V\right)+f\left( g^{\rho\sigma}F^a_{\mu\rho}F^a_{\nu\sigma}-g_{\mu\nu}\frac{1}{4}F^2\right) & =  T_{\mu\nu} \;,
\end{align}
together with Eq.~\eqref{eq:eom_gauge_general} which clearly is not modified by the choice we imposed on $P(X, \phi)$. While these equations hold for both isotropic and anisotropic configurations, in the rest of this appendix we specialize our equations to these two cases respectively.

\subsection{The isotropic case}
\label{appendix:isotropic}
Using the ansatz described in the bullet point~\ref{cases:isotropic} of Sec.~\ref{sec:model_definition} specified by Eq.~\eqref{eq:FLRW} and Eq.~\eqref{eq:isotropicansatz} we can express the components of the Field Strength tensor as:
\begin{equation}
F^a_{00}=0\;, \qquad F^a_{ij}=0\;, \qquad F^a_{0i}=\dot{v}_A\delta^a_i\;,
\end{equation}
where dots denote derivatives with respect to the cosmic time $t$. These equations can be used to show that:
\begin{equation}
F^2=-\frac{6 \dot{v}_A^2}{a^2}\;, \qquad \frac{\partial}{\partial t}F^2=-2F^2\left(H-\frac{\ddot{v}_A}{\dot{v}_A}\right)\;,
\end{equation}
where, in order to ease the notation, we defined $F^2 \equiv g^{\mu\nu} g^{\alpha\beta} F^a_{\mu \alpha}F^a_{\nu \beta} $. At this point we can express the e.o.m. for the inflaton and for the gauge fields as:
\begin{align}
	\ddot{\phi}+3H\dot{\phi}+V_{,\phi} & = \frac{3\dot{v}_A^2}{2 a^2} \left[\ddot{\phi}f_{,X}+Hf_{,X}\dot{\phi} +f_{,X\phi}\dot{\phi}^2 +f_{,XX}\dot{X}\dot{\phi}+2\frac{\ddot{v}_A}{\dot{v}_A}f_{,X}\dot{\phi}+f_{,\phi}\right] \;,\\
	 \partial_{0}\left(a f \dot{v}_A\right) & = 0 \;, \label{eq:gauge_eq_isotropic} 
\end{align}
Moreover, it is possible to show that the only non-vanishing components of the Einstein equations are:
\begin{align}
	3H^2&=T_{00} = \frac{\dot{\phi}^2}{2}\left(1-\frac{3 f_{,X}\dot{v}_A^2}{a^2}\right)+V(\phi)+\frac{3f\dot{v}^2_A}{2a^2}\;,\\
	-\delta_{ij}2\dot{H}&=\frac{T_{ij}}{a^2}+\delta_{ij}T_{00}=\delta_{ij}\left[\dot{\phi}^2\left(1-f_{,X}\frac{3 \dot{v}_A^2}{2a^2}\right)+\frac{2 f\dot{v}_A^2}{a^2}\right]\;.
\end{align}
From the last two equations we can immediately compute the first slow-roll parameter:
\begin{equation}
	\epsilon_H \equiv - \frac{\dot{H}}{H^2} = \frac{3}{2} \frac{\dot{\phi}^2\left[1-3 f_{,X} \dot{v}_A^2 / (2a^2) \right]+ 2 f \dot{v}_A^2/ a^2}{ \dot{\phi}^2 \left[1- 3 f_{,X}  \dot{v}_A^2 / a^2\right] / 2+V(\phi)+ 3f\dot{v}^2_A/(2a^2)}  \; ,
\end{equation}
from which we can immediately conclude that, if $f >0 $ (implying a positive kinetic term for the gauge fields) and $f_{,X} < 0$ (recall that $X = -\dot{\phi}^2/2 < 0$), a small $\epsilon_H$ (and thus inflation) can only be obtained if the three conditions:
\begin{equation}
\frac{\dot{\phi}^2}{2} \ll V \; , \qquad \frac{3}{2} \frac{X  f_{,X} \dot{v}_A^2}{a^2} \ll V \; , \qquad \frac{f\dot{v}^2_A}{a^2} \ll V \; ,
\end{equation}
are simultaneously satisfied.

\subsection{The anisotropic case}
\label{appendix:anisotropic}
Using the ansatz described in the bullet point~\ref{cases:anisotropic} of Sec.~\ref{sec:model_definition} specified by Eq.~\eqref{eq:anisotropicgauge} and Eq.~\eqref{eq:anisotropicmetric} we can express the components of the Field Strength tensor as:
\begin{equation}
	F_{01}=-F_{10}=\dot{v}_A\;.
\end{equation}
Analogously to the isotropic case this can be used to show that:
\begin{equation}
	F^2=-2\frac{b^4 \dot{v}^2_A}{a^2}\;, \qquad \frac{\partial }{\partial t} F^2=F^2\left(4H_b-2H_a+2\frac{\ddot{v}_A}{\dot{v}_A}\right)\;,
\end{equation}
where we have introduced $H_a\equiv \dot{a}/a$ and $H_b\equiv\dot{b}/b$. The e.o.m. for the inflaton field, for the gauge field and the stress-energy tensor become:
\begin{align}
	\ddot{\phi}+3H_a\dot{\phi}+V_{,\phi}=&-\frac{1}{4}F^2\left[\ddot{\phi}f_{,X}+f_{,\phi} + f_{,X}\dot{\phi} \left(H_a+4H_b + 2\frac{\ddot{v}_A}{\dot{v}_A}\right)  +\right.\notag\\
	&\hspace{1.5cm} \left. + f_{,XX}\dot{X}\dot{\phi} +f_{,X\phi} \, \dot{\phi}^2 \right]\;,\\
 	  \partial_0(ab^4f\dot{v}_A)=&\, 0\;.
\end{align}
Moreover, it is possible to show that the only non-zero components of the Einstein equations in the anisotropic case are:
\begin{eqnarray}
 &  \hspace{-1.7cm} 3\left(H_a^2-H_b^2\right) = T_{00}=\frac{\dot{\phi}^2}{2}\left(1-f_{,X}\frac{b^4}{a^2}\dot{v}^2_A\right)+V(\phi)+\frac{b^4}{2a^2}f\dot{v}^2_A\;,\\
& \hspace{-0.7cm}- \left(3H_a^2+3H_b^2+6H_aH_b+2\dot{H}_a+2\dot{H}_b\right) = \frac{T_{11} b^4}{a^2 }= \left(\frac{\dot{\phi}^2}{2}-V(\phi)-\frac{b^4}{2a^2}f\dot{v}_A^2\right)\;,\\
&\hspace{-1.7cm} - \left(3H_a^2+3H_b^2-3H_aH_b+2\dot{H}_a-\dot{H}_b\right) = \frac{T_{22} }{a^2 b^2 }= \frac{T_{33} }{a^2 b^2 } = \left(\frac{\dot{\phi}^2}{2}-V(\phi)+\frac{b^4}{2a^2}f\dot{v}_A^2\right)\;.
\end{eqnarray}
This system can be rearranged to:
\begin{align}
	3\left(H_a^2-H_b^2\right)&=\frac{\dot{\phi}^2}{2}\left(1-f_{,X}\frac{b^4}{a^2}\dot{v}^2_A\right)+V(\phi)+\frac{b^4}{2a^2}f\dot{v}^2_A\;,\label{eq:full_anisofried1}\\
	3H_aH_b+\dot{H}_b&=\frac{b^4}{3a^2}f\dot{v}_A^2\;,\label{eq:full_anisofried2}\\
	6H_a^2+2\dot{H}_a&=-\frac{\dot{\phi}^2}{2}f_{,X}\frac{b^4}{a^2}\dot{v}^2_A+2V(\phi)+\frac{b^4}{3a^2}f\dot{v}_A^2\;.\label{eq:full_anisofried3}
\end{align}
Plugging the first of these equations into the last one and dividing by $H_a^2$ we can obtain the analogous of the first slow-roll parameter:
\begin{equation}
- \frac{\dot{H}_a}{H_a^2} =  \frac{ \dot{\phi}^2 }{2  H_a^2} \left( 1 - \frac{1}{2} f_{,X}\frac{b^4}{a^2}\dot{v}^2_A\right)  + 3 \left( \frac{H_b}{H_a}\right)^2  + \frac{b^4}{3a^2} \frac{f\dot{v}_A^2}{H_a^2}   \;,
\end{equation}
since $f > 0$ (to avoid a negative kinetic term for the gauge fields) and $f_X < 0$ this is a sum of positive terms. As a consequence, in order for inflation to be realized, all of this contributions must be much smaller than one.

\subsubsection{Anisotropic superpotential and useful quantities}
\label{appendix:some_more_eqs}
In this section we report some additional equations which are useful in order to study inflationary models in terms of the $\beta$-function formalism. First of all, from the definition of $\beta_0$ and $\gamma$ we can immediately compute the expression of the two superpotentials $W_a$ and $W_b$:
\begin{equation}
 W_a(\phi) = W_{a,f} \exp\left\{-\frac{1}{2}\int_{\phi_f}^\phi   \textrm{d} \phi' \beta_0(\phi')\right\}\;, \qquad \qquad W_b(\phi) = \gamma(\phi) W_a(\phi)\;,
\end{equation}
where $W_{a,f}$, $W_{b,f}$ and $\phi_f$ denote the values of $W_{a}$, $W_{b}$ and $\phi$ at the end of inflation respectively. From the definition of $\beta$ (and again using the definition of $\gamma$) we can then compute the expression of the two scale factors $a$ and $b$:
\begin{equation}
a(\phi) = a_f \exp\left\{\int_{\phi_f}^\phi  \textrm{d}\phi' \beta^{-1}(\phi')\right\} \;, \qquad b(\phi) = b_f \exp\left\{\int_{\phi_f}^\phi \textrm{d}\phi' \gamma(\phi')\beta^{-1}(\phi')\right\} \;,
\end{equation}
where $a_f$ and $b_f$ are the values of $a$, $b$ at the end of inflation. \\

From the definition of $\gamma$ it is then possible to show that:
\begin{align}
\label{eq:gamma_phi}
\gamma_{,\phi} & = \gamma \frac{W_{b, \phi} }{W_b }  - \frac{W_{a, \phi} }{W_a } \gamma = \gamma \left(  \frac{W_{b, \phi} }{W_b } + \frac{\beta_0 }{2} \right) \;,
\end{align}
which, together with Eq.~\eqref{eq:anisofried2} and Eq.~\eqref{eq:anisofried3}, can be used to compute the expression of the inflaton potential:
\begin{equation}
V\left(\phi\right)=\frac{3}{4} W_a^2\left\{1- \frac{\gamma}{2} - \frac{\beta}{3}\left[\frac{\beta_0}{2} - \frac{\gamma}{2} \left(\frac{\beta_0}{2}- \frac{\textrm{d}\, \ln \gamma}{\textrm{d}\, \phi}\right)\right] \right\}\; .
\label{inflpotbeta}
\end{equation}
Substituting into Eq.~\eqref{eq:anisofried1} and the gauge field energy contribution:
\begin{equation}
\frac{2 \rho_A}{W_a^2 } = \frac{p_A^2}{\tilde{f}\, W_a^2 }=\frac{9}{4} \gamma\left[1- \frac{\beta}{3}\left(\frac{\beta_0}{2}- \frac{\textrm{d}\, \ln \gamma}{\textrm{d}\, \phi}\right)\right] \; ,
\label{gaugeenergybeta}
\end{equation}
which can directly be used to compute $\rho_A$ defined in Eq.~\eqref{eq:rho_A_def}.

\section{Specific examples}
\label{appendix:examples}
We present some explicit and worked out examples of the use of the $\beta$-function formalism on both isotropic and anisotropic cases. In particular we illustrate how the consistency conditions derived for both cases are helpful in model building. Note that these examples have been chosen for illustrating purpose. We are not focusing in their phenomenological prediction and consider them as toy models.

\subsection{Isotropic inflation - Chaotic example}
\label{sec:isotropic_example}
The model being completely defined by $\beta_0$ and $\tilde{f}$, we first pick a $\beta$-function and then show how does constrain the possible parametrization  of $\tilde{f}$. The usual choices for $\beta_0$ are defined in~\cite{Binetruy:2014zya}. For this example we decide to work with the chaotic class:
\begin{align}
	\beta_0 &= - \frac{\alpha}{\phi} \;,
\end{align}  
where $\alpha$ is a positive constant. By definition, the end of inflation occurs at $\epsilon_H = 1$ and we obtain an equation the value of $\phi$ at the end of inflation, $\phi_f$:
\begin{align}
	\beta_0^2\left(4p_A^2-\tilde{f}W_f^2\right)+2\tilde{f}W_f^2&=0\;.
\end{align}
The superpotential and potential for this kind of model are:
\begin{align}
	 W = W_f \left( \frac{\phi}{\phi_f} \right)^{\alpha} \; , && V(\phi) \simeq \frac{3}{4} W_f^2 \left( \frac{\phi}{\phi_f} \right)^{2 \alpha}  -\frac{3p_A^2}{2\tilde{f}(\phi)}\;.
\end{align}  
The  parametrization for $\tilde{f}$  consistent with starting with a nearly dS universe (which is realized for $\beta_0 \rightarrow 0$) are constrained by the condition  Eq.~\eqref{eq:condition_isotropic}. In this case, the condition reads:
\begin{align}
	\beta_{0}^2 &\geq \frac{32p^2_A}{\tilde{f} W^2}  \;, &&  \text{or} && \tilde{f} \geq \frac{32p^2_A}{ W_f^2 \alpha^2 } \left( \frac{\phi}{\phi_f} \right)^{-2 \alpha}  \phi^{2} \; .
	\label{eq:condisochao}
\end{align}
Two simple parametrizations are
	\begin{itemize}
		\item A strict proportionality in~\eqref{eq:condisochao}:
		\begin{equation}
		\tilde{f} = C \frac{32p^2_A}{ W_f^2 \alpha^2 } \left( \frac{\phi}{\phi_f} \right)^{-2 \alpha}  \phi^{2} \;,
		\end{equation} 
		with $C \geq 1 $. In this case the square root in~\eqref{eq:dot_phi_isotropic} is simply constant and the most of the computation are analytical.
		\item A more general choice:
		\begin{equation}
		\tilde{f} = g(\phi) \frac{32p^2_A}{ W_f^2 \alpha^2 } \left( \frac{\phi}{\phi_f} \right)^{-2 \alpha}  \phi^{2}\;,
		\end{equation}
		where $g(\phi) \geq 1$ is a generic function of $\phi$.
	\end{itemize} 
Since isotropic inflation is not the main focus of this work we do not include the full analytical study of these toy models.
 Note that we could also have started with a choice of $\tilde{f}$ and deduced the allowed $\beta$-function from the consistency condition.

\subsection{Anisotropic examples}
\label{sec:appendix_anisotropic}
To describe other concrete applications of the formalism, we present a few possible parametrizations for the function $\gamma$ (alternative to the one used in Sec.~\ref{sec:anisotropic_chaotic}) that satisfy all the consistency conditions for the anisotropic case (\emph{i.e.} Eq.~\eqref{eq:anisobetaconditions1} and Eq.~\eqref{eq:anisobetaconditions2} ).

\subsubsection{Power-law inflation}
Let consider the simplest choice for  $\beta_0$ and $\gamma$ where both of them are constant:
\begin{align}
\beta_0(\phi) &= - C_1 \;, && \gamma(\phi) = C_2\;,
\end{align}
with $C_1$, $C_2$ positive constants. It is easy to show that in this case $\Delta$ is a constant:
\begin{align}
\Delta & = 1 - \frac{ 24 \, C_2}{(1+C_2) C_1^2}  \; , 
\end{align}
and the conditions~\eqref{eq:anisobetaconditions1} and~\eqref{eq:anisobetaconditions2} constraining the parameters $C_1$ and $C_2$ are trivially satisfied by any choice with $C_1, C_2\ll 1$ and $24 C_2 \leq C_1^2$. This very simple parametrization corresponds to the anisotropic power-law model of inflation discussed in~\cite{Kanno:2010nr}. This is made explicit considering the potential:
\begin{align}
V(\phi) & = V_f\exp\left\{{C_1}(\phi-\phi_f)\right\}\;,
\end{align}
where $V_{f}$ is combination of $W_f$, $C_1$ and $C_2$, and the coupling $f = a^{-4}b^{-4}\tilde{f}$:
\begin{align}
f(\phi) &= f_f \exp\left\{\left(-C_1 + \frac{8}{C_1(1+\sqrt{\Delta}) }\right)(\phi-\phi_f)\right\}\; .
\end{align}
where $f_{f}$ is combination of $W_f$, $p_A$ $C_1$ and $C_2$. In particular, by comparing with the model of~\cite{Kanno:2010nr} we easily find that:
\begin{align}
\lambda & = C_1 \;, && \rho = -C_1 + \frac{8}{C_1(1+\sqrt{\Delta})}\;,
\end{align}
and the condition found on the parameters $\lambda$ and $\rho$ (\emph{i.e.} Eq.~32 of~\cite{Kanno:2010nr}) corresponds to:
 \begin{equation}
 	\frac{3 -  \sqrt{\Delta}}{1 + \sqrt{\Delta}} > \frac{C_1^2}{4} \; ,
 \end{equation}
implying for example $C_1<2$ for $\Delta = 1$, which is automatically satisfied here. In this simple case, since $\beta_0$ and $\beta$ are constant, so it is $\epsilon$ and the period of inflation will never end. However this formalism is well suited to construct a better defined model, by adding a correction to the $\beta$ function, as in~\cite{Cicciarella:2017nls}. A choice of $\beta_0= - C_1- F(\phi)$ (where $F(\phi) > 0$ is a generic function which goes to zero for large values of $\phi$) and $\gamma = C_2$ will have most of the features of the strictly anisotropic power-law model but will allow inflation to end.

\subsubsection{Logarithmic parameterization}
Another elegant Ansatz for $\beta_0$ and $\gamma$ is:
\begin{align}
\gamma_{,\phi}&=\frac{1}{2}\beta_0\gamma\;,
\end{align}
which gives:
\begin{align}
\gamma(\phi) &= \gamma_f\exp\left\{\frac{1}{2}\int_{\phi_f}^\phi d\phi'\beta_0(\phi')\right\} = \frac{C}{W_a}\;,
\end{align}
where $C \equiv \gamma_f  W_{a,f}$ is a dimensionful constant. Note that by definition of $\gamma$, we find $W_b=C$ to be constant. The other relevant quantities are given by simple expressions:
\begin{align}
\beta(\phi)&= \frac{1}{2}\beta_0(\phi)\left(1+\sqrt{1-\frac{24\gamma(1+\gamma)}{\beta_0^2}}\right)\;,\\
\tilde{f}&=\frac{4p_A^2}{9 C W_a} \, \;,\\
V(\phi)&=\frac{3W_a^2}{4}\left[1-\frac{\gamma}{2}-\frac{1}{6}\beta\beta_0\right].
\end{align}
The conditions to realize inflation in the model simply reduce to $\gamma \ll1$ and $ \beta_0^2\ll1$ and for the square root to be well defined we obtain:
\begin{align}
\beta_0^2 > 24\gamma(1+\gamma)\;.
\end{align}
This model can be directly computed for any of the known class of universality upon some validity check given by the above consistency conditions. To proceed further and illustrate the mechanism in greater details, let us pick a explicit choice for $\beta_0$, namely the function of the chaotic class $\beta_0 = -\alpha / \phi$, where $\alpha$ is a constant. This choice directly gives:
\begin{align}
\gamma(\phi) & = C_1 \phi^{-\alpha/2}\;, && \beta(\phi)  = - \frac{\alpha}{ 2 \phi}\left(1+\sqrt{1-\frac{24C_1}{\alpha^2}\phi^{\frac{4-\alpha}{2}}}\right)\;,
\end{align}
where $C_1$ is a constant and the condition from the square root gives:
\begin{align}
\phi^{\frac{\alpha-4}{2}} > \frac{24 C_1^2}{ \alpha^2}\;,
\end{align}
implying that $\alpha \geq 4$ and an appropriate choice of $C_1$ should satisfy this condition.  We can not proceed further without fixing $\alpha$. For example, the case with $\alpha = 4$ gives $\beta = - C_2 / \phi$ where $C_2\equiv 2\left(1+\sqrt{1-3C_1/2}\right)$ is a constant. We can compute the coupling between the gauge and the scalar fields:
\begin{align}
f(\phi) & \simeq f_f \exp \left\{  \frac{2}{C_1} \left( \phi^2 - \phi_f^2 \right) \right\} \left(\frac{\phi}{\phi_f}\right)^{4 \left( \frac{C_1}{C_2} - \frac{1}{2}\right) }\;,
\end{align}
where $f_f$ is a constant factor. As a consequence $f$ has to decrease exponentially during inflation (\emph{i.e.} as $\phi$ decreases towards $\phi_f$).

\newpage 

\bibliographystyle{hunsrt}
\bibliography{article.bib}

\end{document}